\documentclass[letterpaper]{article} % DO NOT CHANGE THIS
\usepackage{aaai25}  % DO NOT CHANGE THIS
\usepackage{times}  % DO NOT CHANGE THIS
\usepackage{helvet}  % DO NOT CHANGE THIS
\usepackage{courier}  % DO NOT CHANGE THIS
\usepackage[hyphens]{url}  % DO NOT CHANGE THIS
\usepackage{graphicx} % DO NOT CHANGE THIS
\usepackage{natbib}  % DO NOT CHANGE THIS AND DO NOT ADD ANY OPTIONS TO IT
\usepackage{caption} % DO NOT CHANGE THIS AND DO NOT ADD ANY OPTIONS TO IT
\usepackage{multirow}
\usepackage{graphicx}
\usepackage{float}
\usepackage{arydshln} 
\usepackage{algorithm}
\usepackage{algorithmic}
\usepackage{graphicx} % DO NOT CHANGE THIS
\usepackage{amsmath}
\usepackage{mathrsfs}
\usepackage{amssymb}
\usepackage{svg}
\usepackage{graphicx}
\usepackage{natbib}  % DO NOT CHANGE THIS AND DO NOT ADD ANY OPTIONS TO IT
\usepackage{caption} % DO NOT CHANGE THIS AND DO NOT ADD ANY OPTIONS TO IT
\usepackage{algorithm}
\usepackage{algorithmic}

\usepackage{newfloat}
\usepackage{listings}
\usepackage{newfloat}
\usepackage{listings}
\DeclareCaptionStyle{ruled}{labelfont=normalfont,labelsep=colon,strut=off} % DO NOT CHANGE THIS
\floatstyle{ruled}
\newfloat{listing}{tb}{lst}{}
\floatname{listing}{Listing}
\nocopyright
\title{MRSE: An Efficient Multi-modality Retrieval System for Large Scale E-commerce}
\author{
    Hao Jiang\textsuperscript{\rm 1,3} ,
    Haoxiang Zhang\textsuperscript{\rm 3},
    Qingshan Hou\textsuperscript{\rm 2},
    Chaofeng Chen\textsuperscript{\rm 1}, \\
    Weisi Lin\textsuperscript{\rm 1}\thanks{Corresponding author: wslin@ntu.edu.sg},
    Jingchang Zhang\textsuperscript{\rm 3},
    Annan Wang\textsuperscript{\rm 1}
}
\affiliations{
    \textsuperscript{\rm 1}Nanyang Technological University (NTU), Singapore\\
    \textsuperscript{\rm 2}National University of Singapore (NUS), Singapore\\
    
    \textsuperscript{\rm 3}Shopee, Singapore\\
    \vspace{0.5em} % Add a small vertical space for separation
    \{jianghao907, haoxiang.z.f, houqingshancv, chaofenghust\}@gmail.com, 
}

\usepackage{bibentry}
\usepackage{subfigure}

\begin{document}

\maketitle

\section{Abstract}

Providing high-quality item recall for text queries is crucial for developing search systems in large-scale e-commerce. Current solutions employ Embedding-based Retrieval Systems (ERS) to embed queries and items into a low-dimensional space. However, uni-modality ERS is overly reliant on textual features, rendering it unreliable in complex semantic contexts. Existing multi-modality ERS often neglects individual preferences for different modalities and fails to fully leverage their advantages.
To address these issues and enhance item retrieval performance, we propose MRSE, a Multi-modality Retrieval System for Large Scale E-commerce.
MRSE integrates text, item image, and user preference information through lightweight mixture-of-expert (LMoE) modules, better aligning inter-modality and intra-modality features.
MRSE also generates user portraits on a multi-modality level, capturing user preferences across various modalities.
Additionally, we propose a novel hybrid loss function to enhance multi-modality exploitation, ensuring consistency and robustness through hard negative sampling.
We conduct extensive experiments on a large-scale industrial dataset from the Shopee Search System and online A/B testing, demonstrating the effectiveness of MRSE. The system achieves over 18.9\% offline relevance improvement and 3.7\% online core metrics gain over Shopee’s state-of-the-art uni-modality product understanding system.

% Providing high-quality item recall for text queries is crucial for large-scale e-commerce search systems. Current Embedding-based Retrieval (EBR) solutions embed queries and items into a low-dimensional space. However, uni-modality EBR overly relies on textual features, while multi-modality EBR often neglects individual modal preferences and fails to fully leverage their advantages. We propose MRSE, a Multi-modality Retrieval Search System for Large Scale E-commerce. MRSE integrates text, image, and user preference information through lightweight mixture-of-expert (LMoE) modules, better aligning inter-modality and intra-modality features. It generates user portraits on a multi-modality level, capturing preferences across various modalities. 
% We introduce a novel hybrid loss function with hard negative sampling to enhance multi-modality exploitation, ensuring consistency and robustness. Extensive experiments on Shopee's large-scale industrial dataset and online A/B testing demonstrate MRSE's effectiveness, achieving over 10\% offline relevance improvement and 4\% online core metrics gain over Shopee's state-of-the-art uni-modality product understanding system.
% ---------------------------------------------------------

\begin{figure}[t]
\centering
\includegraphics[width=230pt]{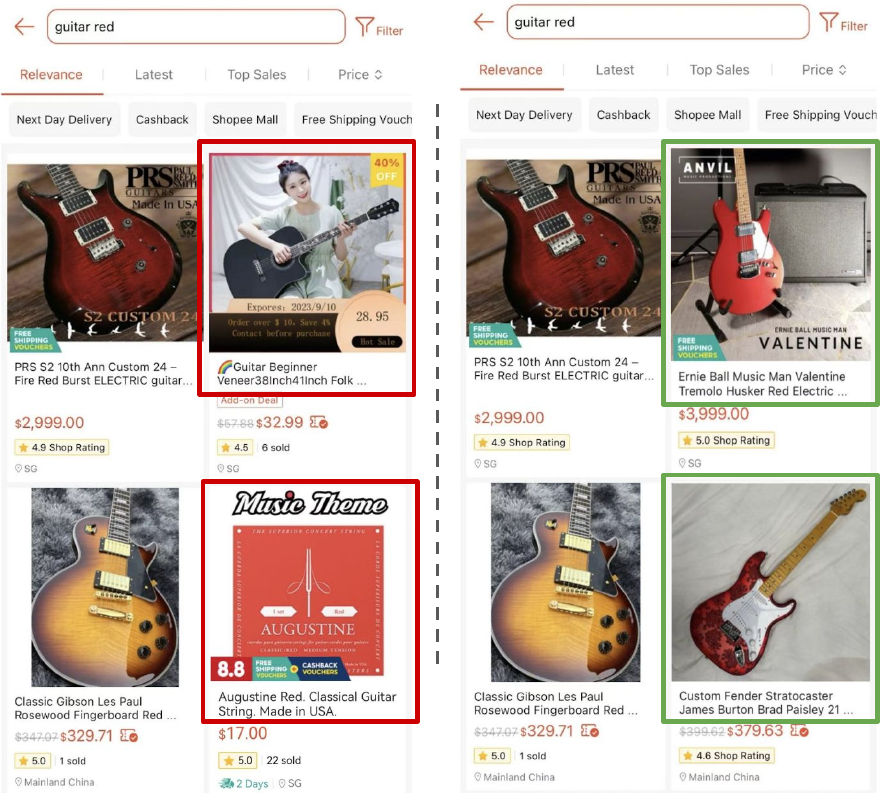}
\caption{Demo of a textual query with color intention: bad cases (left red rectangles) with uni-EBS, and good cases (right green rectangles) with MRSE. The first bad case shows a failure to capture color preference, while the second involves an incorrect match with red guitar strings.}
\label{fig:GBC}
\end{figure}

\section{Introduction}
% Cross-border e-commerce is an essential part of the market, where buyers can purchase products online from different fields, and sellers can save physical rental costs and be impressed by different regions of people. 
% However, cross-border e-commerce faces challenges in improving retrieval performance due to linguistic diversity with multiple granularities, diverse user behavior (i.e., religion, race), and low description quality of item along with unbalanced item pool distribution. 
% % easily scale up the product pool. 
% This study aims to enhance the quality of the Embedding-based Retrieval System (ERS) for product retrieval at Shopee, 
% % and recall of the Marketplace retrieval engine on search, 
% enabling it to recall the most relevant products to users based on their searched textual queries.
Within large-scale e-commerce, multiple scenarios are designed to help users discover items of interest. In the search scenario, users provide textual queries, and the system matches them to items based on their content, as illustrated in Fig.\ref{fig:GBC}.
% illustrates an example of such a query. 
Traditional method \cite{singhal2001modern} uses term match and statistics to build a Retrieval System (RS) based on a large-scale database. 
However, traditional RS makes it hard to handle new and long tail items \cite{yin2012challenging} with the same semantics but different textual terms. 
Subsequently, an Embedding-based Retrieval System (ERS), enriched with semantic depth and operating in high-dimensional spaces has emerged. 
% Several models enable this to happen, 
Word2Vec \cite{mikolov2013efficient} makes a high-level vector space to represent a word, Fasttext \cite{bojanowski2017enriching} moves forward and adopts the n-gram method, building upon the foundation of Word2Vec and enhancing the model's performance, which can be involved in company-level ERS to extract textual embedding with low latency. 
Transformer \cite{vaswani2017attention} uses attention architecture to capture higher-level semantic information in natural language. 
% , which is utilized to capture texual semantic and intention on higher level representation. 
% -------------------------
While these models are highly effective at extracting complex textual semantics, their utility in industrial contexts is constrained by significant discrepancies in textual conventions and features between user queries and item descriptions.
% -------------------------

To improve ERS's capacity for handling diverse features and boosting efficiency, 
% With the advancement of artificial intelligence technology, 
% ERS has been blooming fast over the past decade and has also been utilized in the E-commerce setting. 
Taobao \cite{li2021embedding}, Amazon \cite{nigam2019semantic}, and JD \cite{zhang2020towards} constructed their two-tower ERS: the query tower is used to compute user-oriented features (textual query, behavior, etc.), while the item tower is employed to calculate product-oriented features (item title, description, etc.). However, uni-modality ERS relies entirely on textual information, potentially failing to accurately capture user preferences and misinterpreting text data,  as shown in Fig.\ref{fig:GBC}. Since most items in recommendation systems are presented to users via images, the alignment between images and user queries often directly impacts the probability of clicks. To leverage this, multi-modality fusion of images and text has become a new trend. From early methods using simple weighted sums  
\cite{hidasi2016parallel, liu2021que2search} to recent approaches involving MLP \cite{liang2023mmmlp}, self-attention mechanisms \cite{pan2022multimodal, bian2023multi, zhang2023beyond} and Graph Neural Networks (GNN) \cite{hu2023adaptive}, various fusion methods have emerged, yet they all neglect user preferences for different modality. For instance, a user querying for clothes may focus more on images, whereas a user searching for mops may be more interested in functional descriptions. Constructing effective multimodal retrieval systems that capture the relationship between text queries and image modalities, as well as individual user preferences for different modalities, remains a significant challenge in search systems.

This paper introduces MRSE, an efficient Multi-modality Retrieval System for Large Scale E-commerce. The training of MRSE involves two stages. The first stage has three LMoE modules: Visual-Bert (VBert) to extract inter-modality between image and text, Light-Bert to capture text modality, and Fasttext-Attention (FtAtt) for intra-modality of historical textual features, while the second stage utilizes DSSM to fuse multi-modality representations. In particular, to capture individual modality preferences, we further leverage LMoE modules to create multimodal historical behavior embeddings, allowing for the adjustment of retrieval structures according to modality preferences. Moreover, triplet loss \cite{schroff2015facenet} and softmax cross-entropy loss \cite{li2021embedding} are commonly used in training the ERS model, while it's hard and resource-intensive to achieve global convergence and align multi-modality features during training with billions of data. We further design a hybrid training approach, combining in-batch softmax cross-entropy loss and triplet loss with hard-easy negative sampling, to align multi-modality features more consistently.

The efficacy of MRSE has been validated through online A/B testing and rigorous analysis of a large-scale industrial dataset from Shopee, one of the largest e-commerce platforms in Southeast Asia. As a result of its demonstrated effectiveness, MRSE has been launched as a base model in the overall platform of Shopee. The primary contributions of this work are outlined as follows.
\begin{itemize}
    \item We propose MRSE to utilize query and item image modalities while capturing user preferences across different modalities, delivering an enhanced user-experience-oriented multimodal retrieval.
    \item VBert, FtAtt, and Light-Bert are introduced for better inter-modality and intra-modality leveraging.
    \item A hybrid loss combining in batch softmax cross-entropy loss and triplet loss with customized hard-easy negative sampling to promote more consistent multi-modality alignment. 
    % \item Experiments conducted on a large-scale industrial dataset
    % and online Product Search of Shopee demonstrate the effectiveness of MeSS. Moreover, we do ablation and analyze the effect of each submodel on MeSS.
\end{itemize}

\begin{figure*}[!t]
  \centering
  \includegraphics[width=\textwidth]{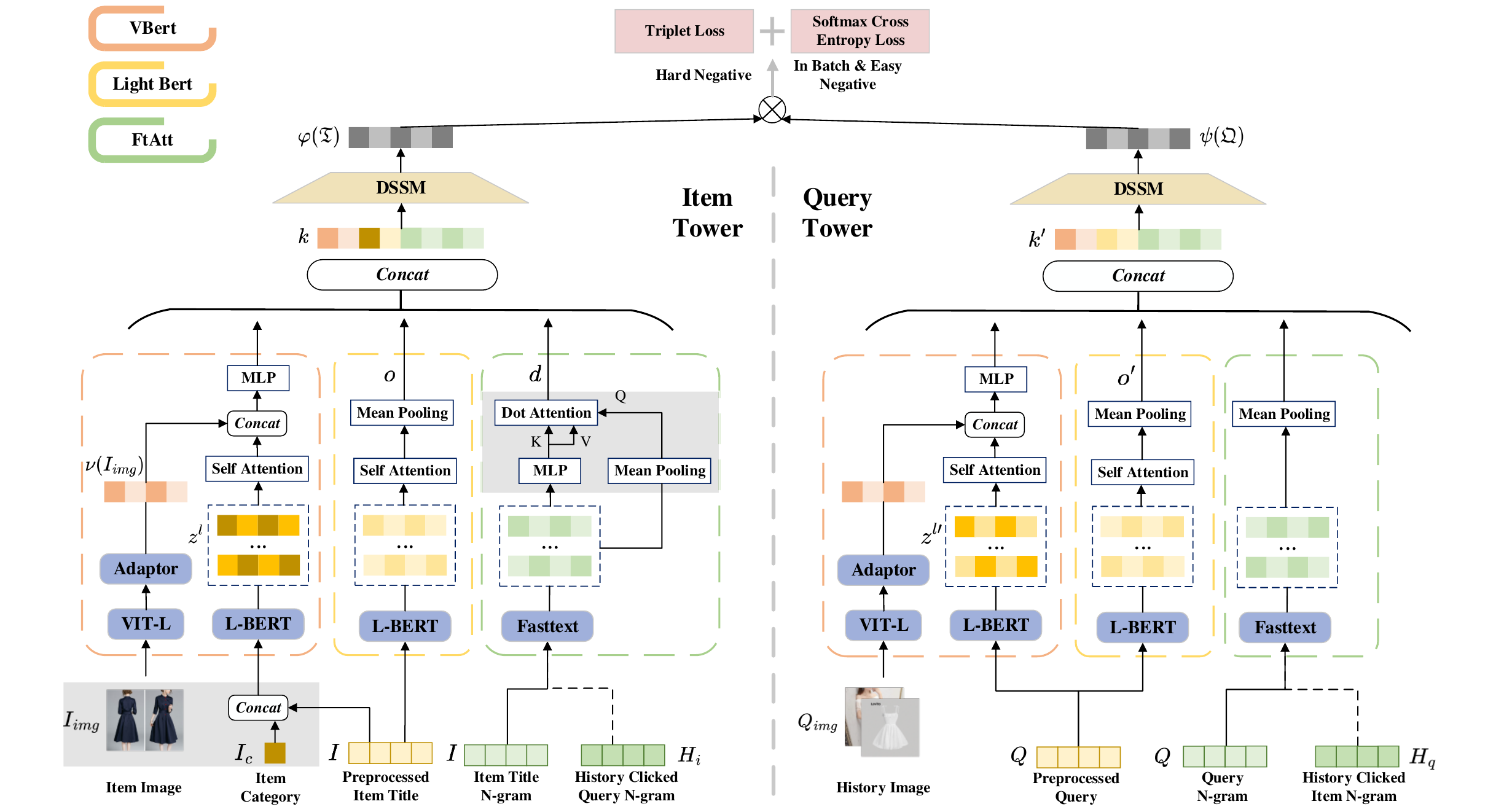}
  \caption{Architecture of MRSE: The multi-modality extraction comprises three LMoE modules, color-coded as shown in the top left corner. In the multi-modality fusion stage, $\varphi(\mathfrak{T})$ and $\psi(\mathfrak{Q})$ represent the final representations for the item tower and query tower, respectively. Key differences between the query and item towers are highlighted with a gray background.}
  \label{fig:overview}
\end{figure*}

\section{Related Work}
\subsection{Matching in Search}
Search matching can be divided into term matching and semantic matching. 
In term matching retrieval, Term Frequency-Inverse Document Frequency (TF-IDF) \cite{ramos2003using} is a widely used technique to represent the importance of terms in a document corpus based on their frequency and rarity. 
% It measures the relevance of a term in a document by weighing its frequency against its occurrence in the entire corpus. 
On the other hand, BM25 (Best Matching 25) \cite{robertson2009probabilistic} is a probabilistic retrieval model that improves upon TF-IDF by considering document length normalization and term saturation effects, resulting in better retrieval performance, particularly for longer documents.
With the booming growth in deep NLP techniques, several neural models have been proposed to tackle the semantic gap and long tail problems raised by traditional term matching during the last few years. 
Word2Vec \cite{mikolov2013efficient}, FastText \cite{bojanowski2017enriching}, and BERT \cite{vaswani2017attention} are commonly applied in semantic matching, employing neural network architectures to generate dense vector representations of words or tokens, facilitating more accurate and general matching in information retrieval.

% BM25 \cite{robertson2009probabilistic} uses 
\subsection{Multi-modality Modeling}
The exploration of multimodal modeling techniques has seen substantial growth over recent years. 
% Due to the diversity of real-world data and the complexity of deep learning applications, multimodal modeling techniques have garnered increasing attention. 
In the Vision and Language domain, Pixel-BERT \cite{huang2020pixel} connects pixel and language modalities within image-text pairs, and UNITER \cite{chen2020uniter} improves multimodal alignment through conditional masking strategies. Similarly, in the context of recommendation systems, diverse fusion methods such as weighted sums \cite{hidasi2016parallel, liu2021que2search}, MLP \cite{liang2023mmmlp, zhao2023m5}, self-attention mechanisms \cite{pan2022multimodal, bian2023multi, zhang2023beyond} and GNN \cite{hu2023adaptive} have been introduced to enhance modality alignment. However, these approaches focus solely on inter-modality and intra-modality features, overlooking the modeling of user modality preferences.

% In recent years, there has been a notable surge in the exploration of multi-modality modeling techniques. Various approaches \cite{dosovitskiy2020image, chen2020uniter, huang2020pixel} have been proposed to effectively integrate text and image modalities. While these text-image interaction techniques have shown exceptional performance when trained on a single language with sufficient data, they are not as effective for low-resource languages due to the lack of adequate training data and pre-train model. In this paper, we propose VBert that builds upon the foundation of text-image interaction by incorporating user query history and query clicked product information to enhance performance.

\subsection{Embedding-based Retrieval in Industry}
% In the e-commerce domain, ERS methods have been employed to improve product search performance while maintaining low latency.
With billions of requests daily in a real-world environment, ERS often utilizes Approximate Nearest Neighbor (ANN) algorithms, such as HNSW \cite{malkov2018efficient}, to efficiently retrieve relevant products based on their embeddings in low latency. To further reduce latency, ERS uses a two-tower architecture to perform late fusion of user query and product information and employs Deep Structured Semantic Models (DSSM) \cite{huang2013learning} to reduce dimensionality. In this paper, we first train the two-tower structure of the LMoE modules and then utilize a DSSM architecture to train the feature fusion model, aiming to optimally balance performance and efficiency.

% In this paper, in the first stage, we introduce the feature encode model VBert, FtAtt, and Light-Bert to maximally balance the performance and latency. 
% % and real-time image-to-image retrieval techniques. To further reduce the latency of the model, we employ lightweight BERT and FtAtt.
% In the second stage, the feature fusion model is trained using a DSSM architecture with a hybrid approach combining in-batch softmax cross-entropy loss and triplet loss with hard-easy negative sampling. This approach aims to enhance the relevance of product search results while maintaining the efficiency benefits of ERS in a large-scale e-commerce setting.

\section{Methodology}
This section introduces MRSE, a Multi-modality Embedding-based Retrieval System at Shopee. We begin by discussing the multi-modal input features and presenting the model architecture. Subsequently, we elaborate on the MRSE training process with a hybrid loss. 
% , which involves employing in-batch softmax cross-entropy loss and triplet loss with hard-easy negative sampling. 

% various loss functions and the easy-hard negative sampling method.

\subsection{Overall architecture}

The general structure of MRSE is illustrated in Fig.\ref{fig:overview}, a two-tower structure that operates in two training stages. 
The query tower takes into account the user's textual query $Q = \{q_1, \ldots, q_u\}$, where $q_u$ denotes a query $q$ searched by user $u$. 
Each $q$ is associated with a corresponding sequence of historically clicked items $H_q = \{i_1, \ldots, i_h\}$ and their respective clicked item images $Q_{img} = \{v_1, \ldots, v_h\}$. 
The item tower uses the item title $I = \{i_1, \ldots, i_n\}$, where $n$ denotes the item pool count, its associated item category $I_c = \{c_1, \ldots, c_n\}$, and the corresponding product image $I_{img} = \{v_1, \ldots, v_n\}$ as input.
Additionally, similar to the query tower, we incorporate item history information, which consists of items clicked by historical queries. For each item $i_n$, we represent this information as $H_i = \{q_1, \ldots, q_h\}$, serving as a sequence input to the item tower.

We now define the task. 
Given the item and query features $Q$, $H_q$, $Q_{img}$, $I$, $I_c$, $H_i$, and $I_{img}$, we aim to return a set of relevant items $I$. We use $\mathfrak{Q}$ to represent the query feature set ($Q, H_q, Q_{img}$) and $\mathfrak{T}$ to represent the item feature set ($I, I_c, H_i, I_{img}$). The query-driven process can be formally defined as:
\begin{equation}
z_{Q2I} = S(\varphi(\mathfrak{T}), \psi(\mathfrak{Q})),
\end{equation}
where $S(\cdot)$ denotes the cosine similarity scoring function, $\varphi(\cdot)$ and $\psi(\cdot)$ are encoders for item and query.
% to produce a 128-dimensional vector, respectively.

% We then can use $\psi(R)$ as the item encoder and $U_{img}$ as the image encoder.

% The user-driven process can be mathematically described as:
% \begin{align}
% z_{I2I} &= F(\psi(R), \psi(\mathfrak{I})) + F(\nu(U_{img}), \nu(I_{img})),
% \end{align}
% where $\psi(R)$ represents the encoded user behavior feature and $\nu(\cdot)$ stands for the image encoder. The detail logics is in MeSS Online Implementation section. 

% \subsection{Query Tower}

\subsection{Satge 1: Multi-modality Modules - LMoE}
% The quality of seller-provided product descriptions in e-commerce marketplaces varies considerably. In numerous instances, attributes of marketplace products are either missing or contain misspellings. 

To address the limitations of the text uni-modality and enhance the robustness of ERS in large-scale scenarios, we propose a lightweight mixture-of-experts (LMoE) approach. This approach incorporates three novel expert modules designed to efficiently process and integrate information from various modalities. 
% Specifically, VBert enhances the capture of inter-modality features, while FtAtt leverages historical data to strengthen intra-modality representation. 
By utilizing these multimodal experts, our method aims to improve the overall performance and reliability of EBS in real-world e-commerce applications.

% This approach utilizes modules capable of extracting query and item information from images and item titles modalities. In our paper, we introduce three novel expert modules designed to effectively process and integrate information from these diverse modalities. By leveraging these multi-modal experts, our method aims to improve the overall performance and reliability of EBS in real-world e-commerce applications where data quality can vary substantially.

% % On query tower, we use three input features, query $Q$,  history features $H_q$ and $Q_{img}$. For Ftatt part, we use n-gram as tokenizer, for VBert and light-Bert, they share a same tokenizer - sentencepiece (spm), which we trained by countries for remain the language 
% The query tower incorporates three input features: the query $Q$, history features $H_q$, and their corresponding image features $Q_{img}$. 
% % Using different tokenizers: 
% % To preprocess these features, we employ different tokenizers depending on the submodel. For the FtAtt submodel, we utilize an n-gram tokenizer, while for VBert and Light-Bert, we use a shared tokenizer called SentencePiece (SPM) \cite{kudo2018sentencepiece}. The SPM tokenizer is trained on country-specific data to preserve the peculiarities of each language maximally. After tokenization, we obtain ($Q_{spm}$, $I_{spm}$) and ($Q_{ngram}$, $I_{ngram}$) as the tokenized representations of the query-item pair. 
% In the following subsections, we present a detailed description of our submodels: VBert, FtAtt, and light-Bert.

\subsubsection{Image Modality - VBert}

The architecture of VBert, illustrated in Fig.\ref{fig:overview}, operates on both query and item towers on MRSE and consists of two major levels. VBert has query and item towers shared with the same structure but different weights. 
The query tower extracts a representation embedding from the user portrait (including user query and user behavior), while the item tower derives embeddings at the item level (item title, item image, and item category).

% In the data preprocessing stage, we tokenize the query and item features using a shared SentencePiece (SPM) tokenizer before feeding the textual inputs into the BERT model. The query tower takes $\mathfrak{Q'} = (Q_{spm}, Q_{img})$ as input, while the item tower uses $\mathfrak{I'} = (I_{spm}, I_c, I_{img})$.
% we also present a lightweight Bert model and a new fasttest\cite{bojanowski2017enriching} model with attention mechanism. Furthermore, we show the model fusion and training strategy of ERS in a large-scale E-commerce scenario. 
The first level of VBert employs a pre-trained VIT-L model \cite{dosovitskiy2020image} as the backbone for image encoding. We introduce an adaptor module \cite{chen2022vision}, consisting of two dense layers, to adapt the VIT-L model to our specific domain. The VIT-L-Adaptor is fine-tuned using item image-image pair from Shopee Image Search, while keeping the VIT-L model frozen. 
The adaptor weights are shared between the query and item towers. 
In the second level, we freeze the fine-tuned VIT-L-Adaptor and train a lightweight BERT model \cite{devlin2018bert} with two layers and a 128-dimensional output for both towers. 

Given the independent semantic in token level on e-commerce, we use token output followed by a customized attention layer as the final representation from Light-Bert, rather than the original pooled output.
\begin{equation}
z^l = \text{Attention}(z^{l-1} + \text{FFN}(\text{MultiHead}(z^{l-1}))), 
\label{eq:bert_layer}
\end{equation}
\begin{equation}
\text{FFN}(x) = \max(0, xW_1 + b_1)W_2 + b_2,
\end{equation}
\begin{equation}
o(z^l) = \frac{1}{n} \sum_{i=1}^n z^l_i, 
\end{equation}
where $z^l$ is the token output from the last layer $l$ of Light-Bert, and $o(z^l)$ denotes the final output of Light-Bert. For the query tower, the initial input $z^0 = (Q)$, while for the item tower $z^0 = (I, I_c)$.
% where $z^l$ is the token output from last layer $l$ of Light-Bert, $\text{d}(z^l)$ is the final output of Light-Bert. For the query tower, $z^0 = (Q, Q_{img})$, and for item tower, $z^0 = (I, I_c, I_{img})$. 
% % from the first token $[CLS]$. 
In the final embedding layer, we concatenate the textual and image embeddings and apply two dense layers to capture inter-modality. 
% The last layer is defined as 

% \begin{equation}
% z_VBert = \text{Trasnformer} , 
% \label{eq:bert_layer}
% \end{equation}

% The model is trained using the in-batch softmax cross-entropy loss function, which will be discussed further in the following section.
% \begin{figure}[t]
% \centering
% \includegraphics[width=150pt]{tst.pdf}
% \caption{Visual Bert (VBert), the image modality model in MeSS}
% \label{fig:VBert1}
% \end{figure}

\subsubsection{Text Modality - FtAtt and light Bert}

The architecture of FtAtt, depicted in Fig.\ref{fig:overview}, consists of two levels. In the first level, both the query and item towers share a common traditional FastText model \cite{bojanowski2017enriching}. 
% The FastText model is trained using the concatenated query $Q_{ngram}$ and item title $I_{ngram}$ in the format of "$Q$ space $I$ space $Q$", where "space" represents a whitespace separator.
In the second level, the query tower applies average pooling on the token embeddings obtained from the FastText model. On the other hand, the item tower introduces a customized attention layer to capture the importance of tokens in the item title:
% \begin{equation}
% r_i \sim \text{Bernoulli}(p), 
% \end{equation}
% \begin{equation}
% f_{\{Q,K\}}(x) = (r \odot x)W_{\{Q,K\}} + b_{\{Q,K\}},
% \end{equation}
\begin{equation}
f_{Q,K}(x) = (r \odot x)W_{Q,K} + b_{Q,K},
\end{equation}
\begin{equation}
f_V(x) = (r \odot \frac{1}{n} \sum_{i=1}^n x_i)W_V + b_V,
\end{equation}
\begin{equation}
d(x) = \text{softmax}(f_Q(x)f_K(x)^T) f_V(x), 
\end{equation}
% where $r$ is a vector of independent Bernoulli random variables ($r_i \sim \text{Bernoulli}(p)$ for dropout with drop rate $p$), $W_{\{Q,K,V\}}$ and $b_{\{Q,K,V\}}$ is the weight matrix and bias vector for the transformation, $\mathbf{b}_Q$ is the bias vector for the query transformation.
where $r$ is a vector of independent Bernoulli random variables ($r_i \sim \text{Bernoulli}(p)$ for dropout with drop rate $p$), $W_{Q,K}$, $W_V$ and $b_{Q,K}$, $b_V$ are the weight matrices and bias vectors for the respective transformations, $d(x)$ denotes the final output of FtAtt's item tower.
For the Light-BERT model, we employ a two-layer BERT architecture, mirroring the same Bert-liked structure used in VBert. 
% To maintain consistency, Light-BERT shares the same SentencePiece (SPM) tokenizer as VBert. 

% \begin{figure}[t]
%   \centering
%   \includegraphics[width=200pt]{Ftatt.pdf}
%   \caption{Fasttext Attention (FtAtt), the text modality model in MeSS}
%   \label{fig:Ftatt}
% \end{figure}

\subsection{Stage 2: Multi-modality Fusion - DSSM}

As seen in Fig.\ref{fig:overview}, for both query and item towers, we concatenate multiple embeddings as input to the DSSM. 
These embeddings represented
as $\varepsilon$ in Eq.\ref{MeSS}, include output embeddings $\varepsilon_{vb}$, $\varepsilon_{lb}$, $\varepsilon_{fa}$ and $\varepsilon_{h}$ from VBert, Light-BERT, FtAtt, and an auxiliary embedding derived from applying FtAtt to the query history ($H_q$) and item history ($H_i$) respectively.

\begin{equation}
k = \text{Norm}(\text{concat}(\varepsilon \ldots) W), W \in \mathbb{R}^{\{D, D-128\} \times d}.
\label{MeSS}
\end{equation}

In Eq.\ref{MeSS}, 
% $\varphi_i$ represents different modality representations, and
the weight matrix $W$ has a variable input dimension that depends on whether history information is available. When history is present, $W \in \mathbb{R}^{D \times d}$; otherwise, $W \in \mathbb{R}^{(D-128) \times d}$, reflecting the reduced input dimension.

The inclusion of query and item history embeddings as additional features serves two crucial purposes: capturing the users' historical preferences on item-query pairs and providing complementary information for intra-modality representation. This multi-faceted input allows the DSSM to leverage both current and historical information, potentially leading to more accurate recommendations. The normalization step (Norm) ensures that the final representation $z$ is properly scaled, facilitating stable training and inference.

\subsection{Hybrid Loss Function}

% \begin{figure}[t]
%   \centering
%   \includegraphics[width=250pt]{HybridLoss.pdf}
%   \caption{Illustration of the proposed Hybrid Loss. The blue cube represents the contrastive matrix, the green cube an easy negative sample, and the red cube a hard negative sample. $N$ is the batch size, $e$ the number of easy negative samples, and $h$ the number of hard negative samples.}
%   \label{fig:HybridLoss}
% \end{figure}

\begin{figure}[t]
  \centering
  \includegraphics[width=250pt]{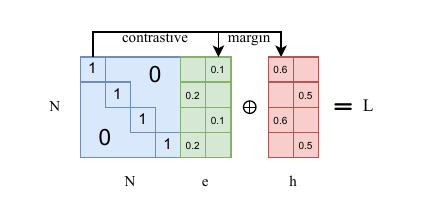}
  \caption{Illustration of the proposed Hybrid Loss. The blue, green, and red cubes represent the contrast matrix, easy negatives, and hard negatives, respectively. $N$ denotes the batch size, $e$ and $h$ are the number of easy negatives and hard negatives.}
  \label{fig:HybridLoss}
\end{figure}

ERS often struggles to simultaneously distinguish highly relevant items from both completely irrelevant ones and somewhat related but non-competitive items. This is particularly problematic for long-tail queries, where false positives can significantly burden subsequent processing stages. 

Traditional two-tower retrieval models often employ the triplet loss function as the training objective. The triplet loss can be defined as:
\begin{equation}
% \hat{y}(q_u, i^+, i^-) = \max(0, \epsilon + F(q_u, i^-) - F(q_u, i^+))
\mathcal{L}(\nabla) = \max(0, \epsilon + S(i^-, \psi(\mathfrak{Q})) - S(i^+, \psi(\mathfrak{Q}))),
\end{equation}
where
% $q_u$ represents the final query representation from the query tower of the model, and 
$i^+$ and $i^-$ denote the final item representations from the item tower of the model. The positive pair $(i^+, \psi(\mathfrak{Q}))$ corresponds to a query-click item pair, while the negative pair $(i^-, \psi(\mathfrak{Q}))$ is formed by the query and a randomly selected item. 
However, the triplet loss often suffers from slow convergence and difficulty in reaching the global minimum. 

% To address this issue, we propose a novel training strategy that combines the softmax cross-entropy loss with both batch-wise and uniformly sampled easy negatives \cite{yang2020mixed}, along with a triplet loss utilizing hard-negative samples derived from the low-relevance logs of the subsequent retrieval stage. This approach is applied to train the VBert and light-BERT models, while the DSSM is trained using only the softmax cross-entropy loss with batch-wise and uniformly sampled easy negatives.

\subsubsection{Softmax Cross-Entropy Loss with mix sampling.}
Building upon Yang et al.'s work \cite{yang2020mixed}, we employ a mix negative sampling softmax cross-entropy loss (MNSCE), defined as:
% The mix sampling softmax cross-entropy loss \cite{yang2020mixed}  is defined as:
\begin{equation}
\hat{y}(i^+| \varphi(\mathfrak{Q})) = \frac{\exp(S(i^+, \psi(\mathfrak{Q}))/\tau)}{\sum_{i' \in I\cup I_e} \exp(S(i', \psi(\mathfrak{Q}))/\tau)},
\end{equation}
\begin{equation}
\mathcal{L}_{mix}(\nabla) = -\sum_{i \in I} y_i \log(\hat{y}_i),
\end{equation}
where $I_e$ is globally random samples from deduplicate item pool, $I$ is the item in current batch, 
$\tau$ is a temperature parameter that controls the smoothness of the overall fitted distribution of the training data. 
As $\tau$ approaches 0, the fitted distribution becomes closer to a one-hot distribution, indicating that the model completely fits the supervisory signals. Conversely, as $\tau$ approaches infinity, the fitted distribution resembles a uniform distribution. 
% In this case, the model is trained to push positive items far away from negative ones, even if the relevance of a positive item is low.
% , suggesting that the model does not fit the supervisory signals at all.
The in-batch softmax cross-entropy loss significantly accelerates the training process. The incorporation of $I_e$ helps alleviate the selection bias inherent in implicit user feedback and mitigates the high resource requirements associated with the in-batch softmax cross-entropy loss, as the resource consumption grows quadratically with the batch size.

\subsubsection{Hard-Negative Triplet Loss}
While MNSCE addresses the long-tail problem to some extent, it overlooks the challenge of hard negatives and fails to sufficiently encourage the model to utilize diverse information from various modalities. To mitigate the hard negative problem and reduce the burden on subsequent stages (such as relevance scoring and ranking), we propose a point-to-point hard-negative sampling triplet loss (HNST) that leverages filtered query-item pairs with low relevance scores obtained from the relevance logs. 
By doing so, we can tackle the hard negative problem without relying on extensive human-labeled data while simultaneously encouraging the model to distinguish between subtly different items across modalities.
The hard-negative triplet loss is defined as:
% The introduction of easy negatives, while beneficial, does not effectively encourage the model to leverage the distinct information provided by different modalities. 
% This limitation is particularly crucial in multi-modal retrieval scenarios, where fully utilizing the complementary nature of various modalities can significantly enhance retrieval performance.

% Despite training on user query-click logs, retrieval models may still encounter false positive items in the recalled set, particularly for long-tail queries. 
% To mitigate the hard negative problem and reduce the burden on subsequent stages (such as relevance scoring and ranking), we propose a point-to-point hard-negative sampling triplet loss (HNST) that leverages filtered query-item pairs with low relevance scores obtained from the relevance logs. 
% By doing so, we can tackle the hard negative problem without relying on extensive human-labeled data while simultaneously encouraging the model to distinguish between subtly different items across modalities.
% The hard-negative triplet loss is defined as:
\begin{equation}
% \hat{y}(q_u, i^+, i_{hard}^-) = \max(0, \epsilon + F(q_u, i_{hard}^-) - F(q_u, i^+))
\mathcal{L}_{hard}(\nabla) = \max(0, \epsilon + S(i_{hard}^-, \psi(\mathfrak{Q})) - S(i^+, \psi(\mathfrak{Q}))),
\end{equation}
where $i_{hard}^-$ represents the representation of the hard-negative item selected from the low-relevance logs.

\subsubsection{Hybrid Loss}
The hybrid loss function combines a weighted sum of MNSCE and HNST, as illustrated in Fig.\ref{fig:HybridLoss}. 
% As in Fig.\ref{fig:HybridLoss}, the blue cube is the contrastive matrix, the green cube is an easy negative, the red cube is the hard negative, N is the batch size, $e$ is the number of easy negative sampling, $h$ is the number of hard negative sampling.
The hybrid loss function is defined as:
% a weighted sum of the softmax cross-entropy loss and the hard-negative triplet loss:
\begin{equation}
\mathcal{L}_{hybrid}(\nabla) = \alpha \mathcal{L}_{mix}(\nabla) + \beta \mathcal{L}_{hard}(\nabla),
\label{fig:hybrid}
\end{equation}
where $\alpha$ and $\beta$ are two hyperparameters that adjust the importance of sample difficulty. As seen in Fig.\ref{fig:modelPerformance}(a), hybrid loss effectively accelerates model convergence and enhances overall model performance. The hybrid approach thus provides a comprehensive solution to leverage different modalities and sample difficulties for EBS in large-scale E-commerce. 
% , selection bias, and long-tail queries in two-tower retrieval models. 
% This approach leads to improve multi-modality feature alignment and reduced computational resource requirements.

\section{MRSE Online Implementation}
\begin{figure}[!t]
\centering
\includegraphics[width=250pt]{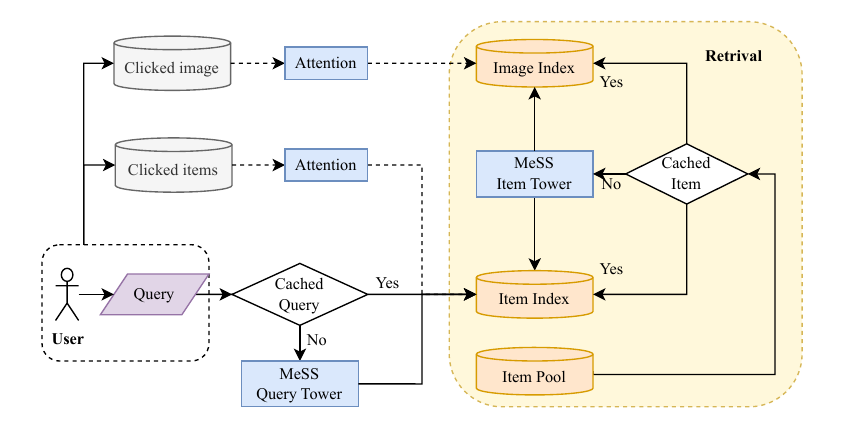}
\caption{Overview of MRSE System Architecture}
\label{fig:System Architecture}
\end{figure}

Fig.\ref{fig:System Architecture} illustrates the online implementation of MRSE. To capture users’ preferences for different modalities and enhance the controllability of the diversity and fairness of retrieval, inspired by \cite{zhang2023divide}, we introduce a `Divide and Conquer' approach to better utilize each modality. This approach incorporates two specialized components: Query-to-Item (Q2I) and Item-to-Item (I2I). The design constructs a dual-item pool index and three recall queues. 
% enhancing retrieval quality and addressing the cold-start problem.

% Fig.\ref{fig:System Architecture} depicts the system architecture of MeSS. 
% In this section, we introduce a novel 'Divide and Conquer' approach for the online implementation of MeSS.
% The MeSS multi-modality system incorporates two specialized queues: Query-to-Item (Q2I) and Item-to-Item (I2I), constructing a dual item pool index and three recall queues to enhance retrieval quality and address the cold-start problem.
% and the online latency of MeSS real-time inference. 

% \subsection{Overall Search System}
% When a user search a query on the Shopee platform, the search engine triggers multiple retrieval queues, i.e, MeSS QI2, MeSS I2I, and term matching. These queues work in parallel to recall a deduplicated, unordered set of the most relevant items, typically numbering in the thousands. The recalled item set then undergoes the relevance and pre-rank stage, which truncates the items to hundreds. Subsequently, the filtered item set passes through the ranking, re-ranking, and mix-ranking stages to determine the final order before being presented to the user as the search results. A noval trade-off between latency and performance is crucial in the overall search system. 
% t is   between latency and accuracy. 

% \subsection{MeSS Online Implementation}

% % \subsection{Real Time User Tower}
% % As shown in Fig.\ref{fig:System Architecture}, 
% During retrieval, we construct a dual item pool index and three recall queues on MeSS to enhance retrieval quality and address the cold-start problem
% % using multi-modality representation user-inquiry (Q2I) and user-behavior (I2I). 
% % This section introduces the online implementation of MeSS for user-inquiry (Q2I) and user-behavior (I2I).
% The MeSS multi-modality system incorporates two specialized queues: a Query-to-Item (Q2I) queue designed to recall items in response to targeted user queries, and an Item-to-Item (I2I) queue engineered to recommend items based on user behavior.

\subsection{MRSE Q2I}

MRSE Q2I is deployed in production at the retrieval stage, operating on products in real time. Model inferences are computed on a cloud of machines called Embed-Producer, with inferences performed by country. To reduce online calculation pressure for user queries, we collect queries from the previous week, pre-calculate the query embeddings, and store them in a query cache. The online query hit rate achieves 90\% for each country.
For item representation, the Embed-Producer pre-calculates the item embeddings to build the MRSE item index, and triggers real-time calculations for new items or when a seller modifies the information of a specific item. The updated embeddings are added to the product index in real time, resulting in an online index hit rate of 99.9\% for each country.

% MeSS Q2I is deployed in production at the retrieval stage and is designed to operate on products created in real time. Model inferences are computed on a cloud of machines called Embed-Producer, with inferences performed by country.
% To reduce online calculation pressure for user queries, we collect queries from the previous week, pre-calculate the query embeddings, and store them in a query cache. 
% % If the query key is not found in the cache, the query is calculated in real-time. The online query hit rate is 90\% for each country.
% % For item representation, 
% The Embed-Producer pre-calculates the item embeddings to build the MeSS item index, 
% and triggers real-time calculations for new items or when a seller modifies the information of a specific item. The updated embeddings are added to the product index in real-time. The online index hit rate is 99.9\% for each country.

% % To enable real-time user behavior retrieval in MeSS, we collect user click history. For MeSS real-time, we maintain a user behavior cache that stores user embeddings based on user clicks with truncation of 20 items. The newly recalled item set is only affected after the front page of the search results.
% % For image user behavior retrieval, we collect user click history with a truncation of 7 items. Due to the resource-intensive nature of real-time image processing at the user level, we calculate the top-100 item-to-item results offline and store them in an image user behavior cache.

\subsection{MRSE I2I}

% the real time user tower incorporates user's real-time click behavior to enhance the retrieval quality and address the cold-start problem. 
We use fine-tuned VIT-L-Adaptor as the image encoder $\nu(\cdot)$ to build the MRSE image index. We utilize Apache Kafka \cite{kreps2011kafka} to stream the user behavior data \lq user clicked items' $R$ and `user clicked images' $U_{img}$ in the cache. When a user searches a query $\mathfrak{Q}$, the cached multi-modality representations under the same query will be triggered and sent to the relevant index. 
The MRSE I2I can be mathematically described as:
% z_{Q2I} = F(\varphi(\mathfrak{T}), \psi(\mathfrak{Q})),
% \begin{equation}
% \begin{split}
% z_{I2I} = S(\mathcal{F}(\varphi(R), \psi(Q)), \varphi(T))\cup \\
%          S(\mathcal{F}(\nu(U_{img}), \psi(Q)), \nu(I_{img})),
% \end{split}
% \end{equation}
% % \begin{align}
% % z_{I2I} = S(\mathcal{F}(\varphi(R), \psi(Q)), \varphi(T)) \cup \\ 
% %           S(\mathcal{F}(\nu(U_{img}), \psi(Q)), \nu(I_{img})),
% % \end{align}
% \begin{equation}
% \mathcal{F}(a,b) = \frac{1}{n} \sum_{i=1}^n (\text{softmax} (b (a \parallel b)^T) (a \parallel b))_i, 
% \end{equation}
\begin{gather}
z_{I2I} = S(\mathcal{F}(\varphi(R), \psi(\mathfrak{Q})), \varphi(\mathfrak{T}))\cup  \notag \\
         S(\mathcal{F}(\nu(U_{img}), \psi(\mathfrak{Q})), \nu(I_{img})), \\[0em]
% \mathcal{F}(a, b) = \frac{1}{n} \sum_{i=1}^n (\text{softmax}(b(a \| b)^T)(a \| b))_i, 
\mathcal{F}(a, b) = \text{softmax}((b(a \| b)^T)(a \| b)), 
\end{gather}
% where $S(\cdot)$ denotes the scoring function, $\varphi(\cdot)$ and $\psi(\cdot)$ are encoders for item and query, respectively.
% where 
% $\mathcal{F}(\cdot)$ represents the attention mechanism, Q is $\mathfrak{Q}$, K and  V is and 
where $\nu(\cdot)$ represents the image encoder VIT-L-Adaptor, $\mathcal{F}(\cdot)$ denotes a customized attention function, $a$ represents the `Query', $a \| b$ represents the `Key' and `Value', and $\|$ is used to denote the concatenation operation.
% where $\nu(\cdot)$ stands for the image encoder VIT-L-Adaptor, $\mathcal{F}(\cdot)$ represents a customized attention funtion, $a$ and $a \| b$ denotes Q and K,V, 
% $\parallel$ represents concatenate function. 
% The detail logics is in MRSE Online Implementation section. 
% allowing for efficient processing and storage of the behavior embeddings in a cache.
% This architecture enables the system to quickly retrieve based on personalized behaviors.
% Within the real-time tower, we employ two personalized item-to-item (i2i) retrieval logics: image-based i2i and vector recall i2i. The image-based i2i leverages the image embeddings obtained from the VIT-L-Adaptor model to find visually similar items. On the other hand, the vector recall i2i utilizes the item embeddings generated from the item tower output of DSSM to identify global semantically similar items.
To strike a balance between efficiency and sufficient behavior information, we truncate the number of cached click behaviors by 20. 
Table.\ref{tab:comparison} demonstrates that incorporating MRSE I2I as an auxiliary component significantly improves recommendation performance. It captures user preferences across various modalities, delivering an enhanced user-oriented multimodal retrieval experience.
% the number of click behaviors considered for each user to a maximum of 20. This truncation helps in managing the memory and processing requirements while still leveraging the most recent and relevant user actions.
% Our A/B test results demonstrate that incorporating the MRSE I2I as an auxiliary component significantly improves the recommendation performance, 
% particularly for new items. By leveraging the visual similarities captured by the image embeddings, the system can effectively address the cold-start problem and provide relevant recommendations for items with limited historical interactions. 

% The integration of real-time user behavior data, along with the dual i2i recommendation logic, enables our system to adapt quickly to user preferences, deliver personalized retrieval items, and handle the challenges associated with cold-start items. 
% The efficient streaming and caching mechanisms ensure that the real-time tower can process and serve recommendations with low latency, enhancing the overall user experience.

\subsection{MRSE Online Latency}
The query embedding is computed within 10 ms at the 99th percentile on CPU at Shopee
scale, while the document embedding is computed within 12 ms. For online indexing, we utilize Vespa to index the item representations by country.

\section{Experiments}
Here we introduce the dataset, configuration of MRSE, evaluation metrics, experimental result, and ablation study to demonstrate the effectiveness of MRSE in the search system. 

\subsection{Dataset}

% We collect user click history data by country from traffic log for 1 week, for SG is 2 weeks due to reletively low traffic data, and filter with spam user. The size of training data varying from 2 billion to 200 million by country. For evaluation, we collect data collect by query, along with the query clicked item. The query set is vary from A to B by country, and the item pool is varying from A to B by country. Same as iamge training data, we collect user click history on image search senario from image traffic log for 1 week, and mix all contries for one dataset because image2image is robust through countries with different languages. 

We collect user click history data as MRSE training data from the traffic log for each country, spanning a period of 1 week. For countries with relatively low traffic, such as Singapore, we extend the data collection to 2 weeks. The collected data is filtered to remove spam users, resulting in a training dataset size that varies from 2k million to 2 billion clicks, depending on the country. 
To evaluate the model performance, we collect query-level data along with the corresponding clicked items by country. The size of the query set ranges from 2 million to 10 million, and the item pool size varies from 600k to 2 million, based on the country. For image training data of VIT-L-Adaptor, we aggregate user click history from the image search scenario traffic log across all countries for a duration of 1 week. This is possible because image-to-image retrieval is robust across countries with different languages, allowing us to train a single image model to serve all countries.

\subsection{Configuration of MRSE}
\begin{table}[t]
% \captionsetup{position=top}
\centering
\resizebox{0.48\textwidth}{!}{%
\begin{tabular}{lcccc}
\hline
Methods & Recall@$K_w^r$ & Recall@$K_u^r$ & Rele@$K_w^p$ \\
\hline
Fasttext & 36.7\% & 42.9\% & 27.6\% \\
Transformer & 46.8\% & 54.6\% & 31.9\% \\
DSSM-$\alpha$ & 54.5\% & 68.6\% & 37.9\% \\
Que2Search & 63.7\% & 79.1\% & 45.8\% \\
\multirow{2}{*}{MRSE Q2I} & 68.2\% & 83.2\% & 48.6\% \\
& (+25.6\%) & (+21.3\%) & (+28.2\%) \\
\textbf{MRSE (Q+I)2I} & \textbf{73.4\%} & \textbf{86.5\%} & \textbf{50.4\%} \\
& \textbf{(+34.7\%)} & \textbf{(+26.0\%)} & \textbf{(+32.9\%)} \\
\hline
\end{tabular}
}
\caption{Comparison with the baseline DSSM-$\alpha$ and multiple baselines on a large-scale industrial offline dataset. We implemented Que2Search as a baseline by incorporating its across-modality attention fusion method and curriculum learning loss into our LMoE framework. (In our experiments, we focus on recalled top-K=500 to align with online performance.)}
\label{tab:comparison}
\end{table}
% For VIT-L-Adaptor, we use two dense layer as adaptor and reduce the output dimension as 128. For VBert and Light-Bert, we use Tiny-Bert with 2 layers and 128 dimension output for both item tower and query tower. For SPM, we set vocabulary size as 30k and truncate by 16 for query tower and 64 for item tower for every Tiny-Bert. For Ftatt, we set ngram as 2 to 4, and output dimention as 256, the we use attention to reduce dimention to 128. On DSSM, we use 1 dense layer and reduce final output to 128 for both item and query tower. 

% As the image motidality extractor of VBert, The 
For the image modality extractor in VBert, the VIT-L-Adaptor employs two dense layers as an adapter, reducing the output dimension to 128. 
For the text modality extractor in VBert and Light-Bert, we utilize a Bert-like structure with 2 layers and output with 128 dimensions for both the item and query towers. SentencePiece (SPM) tokenizer is trained with a vocabulary size of 30k, and truncation is applied with a length of 16 for the query tower and 64 for the item tower for the text modality extractor in VBert and Light-Bert. 
In the FtAtt model, we set the n-gram tokenizer range from 2 to 4 and the output dimension to 256, which is then reduced to 128 using a customized attention mechanism. 
The DSSM consists of two dense layers, reducing the multi-modality representations to 128 for both the item and query towers. 

For hybrid loss on VBert and DSSM, we used $\tau=0.1$ and $\alpha=1$ on MRSE, and $margin=0.15$ and $\beta=30$ on HNST. We used a learning rate of $1e-4$ with a batch size of 256 and Adam optimizer on VIT-L-Adaptor, VBert, and Light-Bert, and a batch size of 1024 on FtAtt and DSSM. 
% and batch size=1024, for softmax cross entropy loss on training VIT-L-Adaptor, VBert, Light-Bert and DSSM, we use $\tau=0.1$ and batch size=1024. 

\subsection{Evaluation Metrics}

% \begin{figure*}[t]
%     \centering
%     \begin{subfigure}
%       \centering
%       \includegraphics[width=100pt]{AnonymousSubmission/LaTeX/figures/Bert Loss.pdf}
%       \caption{Recall@500 of Bert model}
%       \label{fig:Bertmodel}
%     \end{subfigure}
    
%     \begin{subfigure}
%       \centering
%       \includegraphics[width=100pt]{AnonymousSubmission/LaTeX/figures/DSSM Loss.pdf}
%       \caption{Recall@500 of DSSM model}
%       \label{fig:DSSMmodel}
%     \end{subfigure}
% \end{figure*}

\begin{figure}
    \centering
    \subfigure[]{\includegraphics[width=115pt]{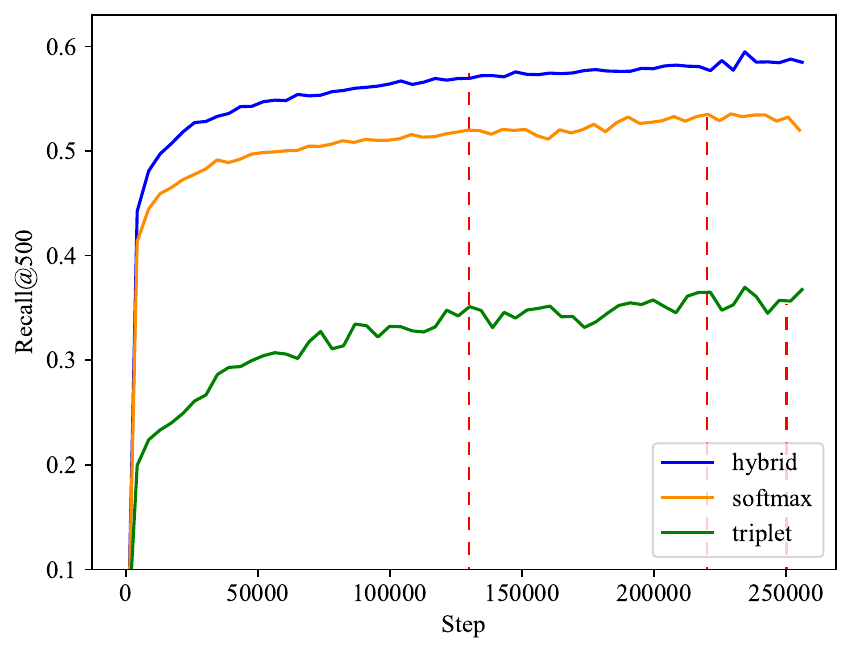}} 
    \subfigure[]{\includegraphics[width=120pt]{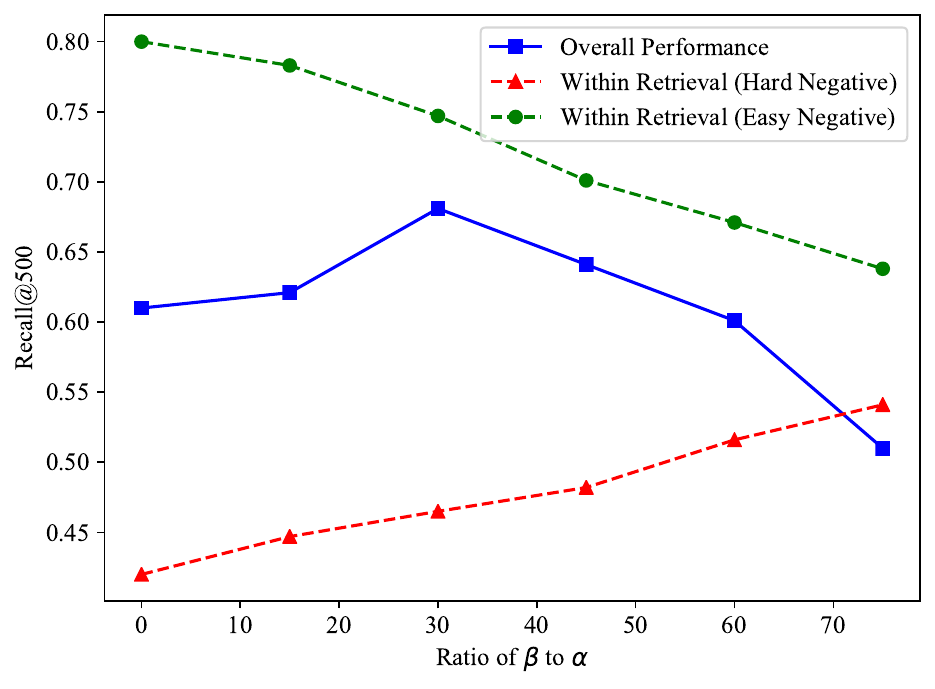}} 
    \caption{(a) Convergence curves of Recall@500 for VBert using triplet loss, softmax loss, and our proposed hybrid loss. (b) The retrieval performance when training MRSE with different ratios of hard negatives rate ($\beta / \alpha$). We conduct two more evaluation datasets with query and its item pool, consisting with the easy negative (randomly picked) and hard negative (from the relevance stage).}
    \label{fig:modelPerformance}
\end{figure}

\subsubsection{Offline Metrics}
We use Recall@$K$ to evaluate the offline performance of MRSE. Specifically, given a query $q_u$, the items clicked or purchased for query $q_u$ are considered as the target set $I_t = \{i_1, \ldots, i_N\}$, while the top-$K$ items returned by the model constitute the retrieval set $I = \{i_1, \ldots, i_K\}$. Recall@$K_q$ is defined as
\begin{equation}
    \text{Recall}@K_q = \frac{\sum_{i=1}^K i_i \in I_t}{N}.
\end{equation}
We introduce Recall@$K_w^r$, Recall@$K_u^r$, and Rele@$K_w^p$ (details in appendix). 

\subsubsection{Online Metrics}
We introduce Impression (IMP), Click-Through Rate (CTR), and Gross Merchandise Volume (GMV) (details in appendix). 

\subsection{Overall Performance}
% As shown in Table.\ref{tab:comparison}, MeSS improves over DSSM-$\alpha$ by 25.6\%, 21.3\%, 36.7\% respectively on Recall@$K_w^r$, Recall@$K_u^r$ and Rele@$K_w^r$, indicating it can retrieve more relevant items within same recalled item set. Rele@$K_u^r$ improves from 37.9\% to 48.6\%, which reduce pressure for relevance stage and ranking stage, and lower the risk for giving a high rank score for a noise but high bidding item. 

% Table \ref{tab:comparison} demonstrates the superior performance of MeSS compared to DSSM-$\alpha$. MeSS achieves improvements of 25.6\%, 21.3\%, and 36.7\% on Recall@$K_w^r$, Recall@$K_u^r$, and Rele@$K_w^r$, respectively, indicating its ability to retrieve more relevant items within the same recalled item set. Moreover, Rele@$K_u^r$ increases from 37.9\% to 48.6\%, reducing the pressure on the relevance and ranking stages and mitigating the risk of assigning high rank scores to noisy but high-bidding items.

To evaluate the effectiveness of MRSE, we conduct a comprehensive comparison against several proven-in-product approaches: Fasttext, Transformer, DSSM-$\alpha$, and Que2Search \cite{liu2021que2search}. To the best of our knowledge, Que2Search is the state-of-the-art multi-modality model for industrial search systems. Specifically, DSSM-$\alpha$ is our previous EBS, which utilizes Transformer and Fasttext as LMoE and DSSM as the decoder.
% and M3SRec \cite{bian2023multi}
% Our previous EBS, DSSM-$\alpha$, utilizes Transformer and Fasttext as encoders and DSSM as the decoder, incorporating statistical data such as sold count and item price in the item tower. 
% Furthermore, we present an ablation study to assess the impact of different modalities and evaluate the importance of each MoE within MeSS. 

Table.\ref{tab:comparison} demonstrates the superior performance of MRSE compared across multiple baselines. Specifically, MRSE Q2I achieves substantial improvements over DSSM-$\alpha$, indicating its ability to retrieve more relevant items within the same recalled item set. Furthermore, MRSE (Q+I)2I shows even greater improvements of 34.7\%, 26.0\%, and 32.9\% on these metrics. A visual comparison between MRSE (Q+I)2I and DSSM-$\alpha$ is illustrated in Fig.\ref{fig:GBC}.
This demonstrates that incorporating multi-modal user behavior enables MRSE to retrieve items based on different modalities, thus adapting to diverse user preferences for various search products. 
Notably, both MRSE Q2I and MRSE (Q+I)2I surpass the performance of Que2Search, indicating that MRSE can better assign and leverage inter and intra modalities information.
% Notably, Rele@$K_w^r$ increases from 37.9\% with DSSM-$\alpha$ to 50.4\% with MeSS (Q+I)2I,
% potentially reducing the computational load on subsequent relevance and ranking stages while mitigating the risk of assigning high rank scores to noisy but high-bidding items.

% \begin{table}[htbp]
% \centering
% \caption{Comparison with the baseline DSSM-$\alpha$ and multiple baselines on a large-scale industrial offline dataset. We implemented Que2Search as a baseline by incorporating its across-modality attention fusion method and curriculum learning loss into our LMoE framework. (In our experiments, we focus on recalled top-K=500 to align with online performance.)}
% \resizebox{0.48\textwidth}{!}{%
% \begin{tabular}{lcccc}
% \hline
% Methods & Recall@$K_w^r$ & Recall@$K_u^r$ & Rele@$K_w^p$ \\
% \hline
% Fasttext & 36.7\% & 42.9\% & 27.6\% \\
% Transformer & 46.8\% & 54.6\% & 31.9\% \\
% DSSM-$\alpha$ & 54.5\% & 68.6\% & 37.9\% \\
% Que2Search & 63.7\% & 79.1\% & 45.8\% \\
% MRSE Q2I & 68.2\% (+25.6\%) & 83.2\% (+21.3\%) & 48.6\% (+28.2\%) \\
% \textbf{MRSE (Q+I)2I} & \textbf{73.4\% (+34.7\%)} & \textbf{86.5\% (+26.0\%)} & \textbf{50.4\% (+32.9\%)} \\
% \hline
% \end{tabular}
% }
% \label{tab:comparison}
% \end{table}
% \usepackage{float} % Add this to your preamble if not already included

\subsection{Ablation Study}

To evaluate the effectiveness of different modalities on MRSE, we conduct an ablation study focusing on Recall@$K$ and Rele@$K_w^p$ ($K=500$). Table.\ref{tab:feature ablation} illustrates the progressive improvements as different features are incorporated. 
MRSE using only SPM and n-gram features achieves a significant improvement on Recall to 60.1\% and Relevance to 41.6\%. A similar impact comes from image features, which we apply in VBert. The inclusion of category features further improves Recall to 63.9\% and Relevance to 45.2\%. Incorporating history features yields a significant improvement, with Recall reaching 68.2\% and Relevance improving to 48.6\%. Finally, the addition of user behavior features results in the best performance, with Recall at 73.4\% and Relevance at 50.4\%.

% Recall of 57.6\% and Relevance of 38.6\%. Adding n-gram features enhances performance, 
% increasing Recall to 60.1\% and Relevance to 41.6\%. The inclusion of category features further improves Recall to 63.9\% and Relevance to 45.2\%. Incorporating history features yields a significant improvement, with Recall reaching 68.2\% and Relevance improving to 48.6\%. Finally, the addition of user behavior features results in the best performance, with Recall at 73.4\% and Relevance at 50.4\%.

These results demonstrate that while pure textual features (SPM and n-gram) provide a strong foundation, the addition of image, categorical, history, and behavioral information substantially enhances the model's capabilities. Notably, the integration of history and user behavior features not only boosts performance (Recall) but also improves the model's robustness and consistency (Relevance) with subsequent search stages, highlighting the importance of multi-modal approaches in e-commerce retrieval systems.

\begin{table}[t]
% \captionsetup{position=top}
\centering
\resizebox{0.48\textwidth}{!}{%
\begin{tabular}{ccc}
\hline
Technique & Recall@$K$ & Rele@$K_w^p$ \\
\hline
SPM feature & 57.6\% & 38.6\%  \\
SPM feature + n-gram feature & 60.1\%& 41.6\%  \\
Above + image feature & 63.3\% & 44.7\% \\
Above + category feature & 63.9\%& 45.2\%  \\
Above + history feature & 68.2\%& 48.6\%  \\
\textbf{Above + user behavior feature} & \textbf{73.4\%} & \textbf{50.4\%}  \\
\hline
\end{tabular}
}
\caption{Ablation study of MRSE: Each modality contributes independent gains. Absent modalities are represented by zero values.}
\label{tab:feature ablation}
\end{table}

\subsection{Importance of submodels}

To evaluate the importance and cooperation of each LMoE modules, we randomly collect 100k queries and items and calculate the cosine similarity between the representations of each modules and the final representations for the item and query towers of MRSE, respectively.
As shown in Table.\ref{tab:model importence}, FtAtt has the highest importance score of 0.637 on the query tower, followed by VBert at 0.613 and Light-Bert at 0.594. This indicates that n-gram features can handle most query cases due to the low semantic knowledge within user queries, while VBert and Light-Bert provide supplementary information for user behavior and intent.
On the item tower, VBert contributes the most with an importance score of 0.682, highlighting the essential role of image features in large-scale e-commerce retrieval. Light-Bert and FtAtt also make significant contributions with scores of 0.585 and 0.536, respectively, suggesting that the item tower relies on deeper LMoE modules for richer semantic information.

% \usepackage{float} % Add this to your preamble if not already included
% \begin{table}[t]
% % \captionsetup{position=top}
% \centering

% % \resizebox{0.33\textwidth}{!}{%
% \begin{tabular}{|c|c|c|}
% \hline
% \textbf{Tower} & \textbf{LMoE} & \textbf{Importance Score} \\
% \hline
% Query & Light-Bert & 0.594  \\
% \hline
% Query & FtAtt & 0.637  \\
% \hline
% Query & VBert & 0.613  \\
% \hline
% Item & Light-Bert & 0.585  \\
% \hline
% Item & FtAtt & 0.536  \\
% \hline
% Item & VBert & 0.682  \\
% \hline
% \end{tabular}
% % }
% \caption{LMoE importance of MRSE.}
% \label{tab:model importence}
% \end{table}

\begin{table}[t]
\centering
% \resizebox{0.48\textwidth}{!}{%
\begin{tabular}{ccc}
\hline
Tower & LMoE & Importance Score \\
\hline
Query & Light-Bert & 0.594  \\
Query & FtAtt & 0.637  \\
Query & VBert & 0.613  \\
% \hdashline
\hline
Item & Light-Bert & 0.585  \\
Item & FtAtt & 0.536  \\
Item & VBert & 0.682  \\
\hline
\end{tabular}
% }
\caption{LMoE importance of MRSE. }
\label{tab:model importence}
\end{table}

% \begin{table}
% \centering
% \begin{tabular}{|l|l|r|}
% \hline
% Tower & Feature & Tower Importance \\
% \hline
% Query & XLM & 60\% \\
% \hline
% Query & Char trigram & 40\% \\
% \hline
% Document & GrokNet & 55\% \\
% \hline
% Document & Description XLM & 13\% \\
% \hline
% Document & Title XLM & 9\% \\
% \hline
% Document & Description Char trigram & 1.5\% \\
% \hline
% Document & Title Char trigram & 1.5\% \\
% \hline
% Document & Language & 1\% \\
% \hline
% Document & Country & 1\% \\
% \hline
% \end{tabular}
% \caption{Feature Importance by Tower}
% \label{tab:feature_importance}
% \end{table}
\subsection{Hybrid Loss Analysis}
% \begin{figure}[!t]
% \centering
% \includegraphics[width=210pt]{ratio.pdf}
% \caption{The retrieval performance when training MRSE with different ratios of hard negatives rate ($\beta / \alpha$). We conduct two more evaluation datasets with query and its item pool, consisting with the easy negative (random picked) and hard negative (from relevance stage). }
% \label{fig:ratio}
% \end{figure}

We analyze the performance of MRSE with different ratios of $\beta$ to $\alpha$. As shown in Fig.\ref{fig:modelPerformance}, as the ratio increases, MRSE gradually improves its ability to distinguish hard negatives. However, this comes at the cost of diminishing performance on easy negatives. Up to a certain ratio, MRSE effectively leverages information from both easy and hard negatives. Beyond this point, the excessive focus on difficult samples leads to a decline in overall performance. 
% We analyze the performance of MRSE with different $\alpha$ and $\beta$. As shown in Fig.\ref{fig:ratio}, with increasing $\beta$, MRSE 
% using hybrid loss, softmax cross-entropy loss, and triplet loss as the training objective, respectively. We report the Recall@$500$ score with respect to the number of training steps on ID (Indonesia), which is the largest market of Shopee. As shown in Fig.\ref{fig:ratio}, the softmax loss's global comparison capability makes faster convergence and better performance. Hybrid further boosts the performance by using mix sampling with multi-loss. 

% We analysis the performance of MRSE using hybrid loss, softmax cross-entropy loss, and triplet loss as the training objective, respectively. We report the Recall@$500$ score with respect to the number of training steps on ID (Indonesia), which is the largest market of Shopee. As shown in Fig.\ref{fig:modelPerformance}, the softmax loss's global comparison capability makes faster convergence and better performance. Hybrid further boosts the performance by using mix sampling with multi-loss. 
% hard engative and easy negative sampling. 
% In fact, it only takes about one 4 hour for the hybrid loss to converge within 4 GPUs while about 1 day for the triplet loss. 
% Note that the margin parameter used in the hinge loss has been carefully tuned. 

\subsection{Online A/Btest}
To assess the real-world impact of MRSE, we conduct an online A/B test on the Shopee platform. Table.\ref{tab:online ab} presents the results, demonstrating improvements of 3.7\%, 3.5\%, and 3.2\% in IMP, CTR, and GMV, respectively, compared to the existing Shopee Search on Ads system. Given Shopee Search's daily volume of billions of transactions, a 3.2\% improvement in GMV translates to tens of millions in additional revenue. This substantial increase demonstrates MRSE's effectiveness in enhancing both the search experience and driving business success.
% These results validate the effectiveness of  MRSE in enhancing the search experience and driving business success.

% Could be used: The comprehensive experimental results and ablation study demonstrate the superiority of MeSS over the baseline DSSM-$\alpha$ model and highlight the significance of incorporating multiple modalities and submodels to improve retrieval performance. The online A/B test further validates the practical impact of MeSS in a real-world e-commerce setting.

\begin{table}[t]
\centering
% \caption{Online A/Btest result of MRSE.}
% \resizebox{0.4\textwidth}{!}{%
\begin{tabular}{cccc}
\hline
Model & IMP & CTR & GMV \\
\hline
Shopee Search on Ads & 3.7\% & 3.5\% & 3.2\%  \\
\hline
\end{tabular}
% }
\caption{Online A/Btest result of MRSE.}
\label{tab:online ab}
\end{table}

\section{Conclusion}
In this paper, we have presented MRSE, a multi-modality embedding-based retrieval system for enhancing search performance in large-scale e-commerce. MRSE addresses challenges of intra and inter modality leveraging in real world through LMoE modules (VBert, FtAtt, Light-Bert) with a hybrid loss approach. The `Divide and Conquer' strategy allows us to extract modality preferences from users' historical behavior, ensuring that the final retrieval truly capitalizes on the multi-modality. Comprehensive experiments demonstrated MRSE's superiority over baselines, with significant offline improvements and online gains in key metrics (IMP, CTR, GMV). The successful deployment of MRSE as a base model in Shopee underscores its effectiveness. We also introduce the online architecture and deployment scheme of our search system to promote community development. 
Industrial multi-modal ERS is still in its early stages.
% necessitating a careful balance between model efficiency and effectiveness. 
In future work, we plan to incorporate user temporal multi-modal information into the search system, such as long-term add-to-cart and order history, to further enhance ERS performance.

\bibliography{aaai25}
\newpage
\clearpage
\end{document}